# A PLUG-AND-PLAY SYNTHETIC DATA DEEP LEARNING FOR UNDERSAMPLED MAGNETIC RESONANCE IMAGE RECONSTRUCTION

*Min Xiao[1], Zi Wang[2], Jiefeng Guo[3], and Xiaobo Qu[2, 1]*

[1]Institute of Artificial Intelligence, Xiamen University, Xiamen, China
[2] Department of Electronic Science, Fujian Provincial Key Laboratory of Plasma and Magnetic Resonance, Xiamen University, Xiamen, China
[3]Department of Microelectronics and Integrated Circuit, Xiamen University, China

## ABSTRACT

Magnetic resonance imaging (MRI) plays an important role in modern medical diagnostic but suffers from prolonged scan time. Current deep learning methods for undersampled MRI reconstruction exhibit good performance in image de-aliasing which can be tailored to the specific k-space undersampling scenario. But it is very troublesome to configure different deep networks when the sampling setting changes. In this work, we propose a deep plug-and-play method for undersampled MRI reconstruction, which effectively adapts to different sampling settings. Specifically, the image de-aliasing prior is first learned by a deep denoiser trained to remove general white Gaussian noise from synthetic data. Then the learned deep denoiser is plugged into an iterative algorithm for image reconstruction. Results on *in vivo* data demonstrate that the proposed method provides nice and robust accelerated image reconstruction performance under different undersampling patterns and sampling rates, both visually and quantitatively.

## 1. INTRODUCTION

Magnetic resonance imaging (MRI) is a non-invasive and non-radioactive diagnostic modality in modern medical imaging techniques [1]. However, due to the k-space undersampling needed for reducing prolonged acquisition time [2-5], MRI image usually contains strong artifacts. Thus, reconstruction method is essential to reconstruct high-quality images from undersampled measurements.

Over the past two decades, many advanced reconstruction methods have been established for undersampled MRI reconstruction, through using different hand-craft priors in image domain [2, 6-9] or k-space domain [5, 10, 11]. To overcome the cumbersome prior determination and performance limitation of these conventional methods, deep learning has become the mainstream in recent works [12-15]. These methods have been demonstrated to achieve state-of-the-art results [16-18], through end-to-end training on the well-curated training dataset. However, considering the complexity of MRI scanning, the collection of large-scale realistic training data remains a big challenge [15, 19].

To address the aforementioned challenge, physics-driven synthetic data learning is an emerging and promising deep learning paradigm for biomedical magnetic resonance [15]. Following magnetic resonance physical models, it can generate massive synthetic data without or with few realistic data for network training, and in particular, its effectiveness has been demonstrated in our focused field of undersampled MRI reconstruction [19]. However, current trained deep learning models are still specific to the sampling settings (e.g., undersampling patterns and sampling rates) used during their training and may cause performance degradation when applied to other tasks under different sampling scenarios.

In this work, we propose a deep plug-and-play iterative reconstruction method for undersampled MRI reconstruction, to effectively adapt to different sampling settings. Specifically, iterations of the plug-and-play method [20] follow steps of the commonly used proximal gradient descent algorithm. Using synthetic data, the plugged image de-aliasing prior is learned by a deep denoiser trained to remove general white Gaussian noise rather than any specific undersampling artifacts. Then the learned deep denoiser is used to replace the proximal operator in the iterative reconstruction process.

Thanks to the fact that the deep denoiser training is independent of the sampling process, the proposed method consistently provides high-quality reconstructed images on *in vivo* data under different undersampling patterns and sampling rates.

## 2. PROPOSED METHOD

### 2.1. Plug-and-play method

MRI reconstruction problem is to recover a desired image $\mathbf{x} \in \mathbb{C}^N$ from the undersampled measured k-space $\mathbf{y} \in \mathbb{C}^M$, and the forward model can be written as

$$\mathbf{y} = \mathbf{A}\mathbf{x} + \boldsymbol{\eta}, \quad (1)$$

where $\mathbf{A} = \mathbf{UFS} \in \mathbb{C}^{M \times N}$ is the forward encoding operator, $\mathbf{S}$ is the coil sensitivity matrix, $\mathbf{F}$ represent Fourier transform, $\mathbf{U}$ is the undersampling operator. $\boldsymbol{\eta} \in \mathbb{C}^M$ is modelled as additive white Gaussian noise.

For this ill-posed problem in Eq. (1), it can be solved using the following optimization model with a regularization term $\mathcal{R}(\cdot)$:

$$\arg\min_{\mathbf{x}} \|\mathbf{y} - \mathbf{A}\mathbf{x}\|_2^2 + \lambda \mathcal{R}(\mathbf{x}), \quad (2)$$

where $\lambda$ is a regularization parameter which is used to balance the data consistency and regularization term. To solve Eq. (2), the proximal gradient descent algorithm is chosen, and its iteration process can be written by [8, 9, 21] as

$$\begin{cases} \mathbf{r}^{(k)} = \mathbf{x}^{(k-1)} + \gamma \mathbf{A}^H (\mathbf{y} - \mathbf{A}\mathbf{x}^{(k-1)}) \\ \mathbf{x}^{(k)} = Prox(\mathbf{r}^{(k)}) \end{cases}, \quad (3)$$

where $k = 1, 2, ..., K$ represents the number of iterations and $\gamma$ is the step size. In Eq. (3), the first sub-equation enforces the k-space data consistency and the second one conducts the proximal operation to achieve image artifacts removal. In plug-and-play methods [20], the proximal operation is usually replaced by a denoiser that removes general white Gaussian noise from the noise-corrupted data.

### 2.2. Deep denoiser

In this work, the proposed plug-and-play method is combined with a deep denoiser to iteratively reconstruct undersampled MRI images using a learned image prior. Specifically, we first train a learnable deep denoiser using paired clean and noisy synthetic data, and then plug it into the iteration reconstruction process. Thus, we re-written Eq. (3) as

$$\begin{cases} \mathbf{r}^{(k)} = \mathbf{x}^{(k-1)} + \gamma \mathbf{A}^H (\mathbf{y} - \mathbf{A}\mathbf{x}^{(k-1)}) \\ \mathbf{x}^{(k)} = DeepDenoiser(\mathbf{r}^{(k)}) \end{cases}. \quad (4)$$

Inspired by the success of 1D learning scheme in undersampled MRI reconstruction [18], we also adopt it to design the network architecture and training strategy of our deep denoiser. The architecture of our denoiser refers to the image de-aliasing module in [19]. Here, we only modify the network's input is noisy image and the output is the denoised image, while other parameters and layers keep the same specifications. Since designed denoiser owns 1D structure, in the iterative reconstruction, we use it to process a 2D image along rows and columns simultaneously. The overall process of our deep plug-and-play method is shown in Figure 1.

To train our deep denoiser, we utilize synthetic data according to [19] rather than scarce realistic data. Different to [19], our data generation is independent to forward encoding process $\mathbf{A}$, and training samples are paired clean and noisy images $(\mathbf{x}_{ref}^t, \mathbf{x}_{noisy}^t)$, where $t = 1, 2, ..., T$ represents the number of training samples. The noisy data are obtained by adding white Gaussian noise with a varied range of signal-to-noise ratio (SNR) on the clean ones.

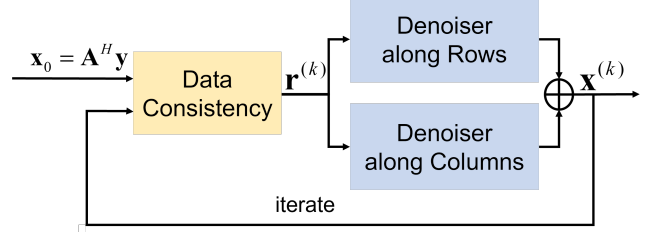

**Fig. 1.** The overall process of our deep plug-and-play method.

### 2.3. Implementation details

Here, we generate paired synthetic training datasets as follows: 1) Select 2000 natural images from ImageNet [22] and convert them to grayscale as image magnitude. 2) Generate random phase through Fourier truncation with size 2~5 and amplitude normalization of the white Gaussian noise [10]. 3) Match above two components randomly and extract 1D images with size 320. 4) Add white Gaussian noise within a range of SNR from 5 to 40 dB randomly, to obtain noisy 1D synthetic data. Thus, 640000 paired 1D synthetic data are generated, and we choose 90% for training and other 10% for validation.

In the training stage, the deep denoiser is trained for 200 epochs with the Adam optimizer. Batch size is 128, and its initial learning rate is set to 0.001 with an exponential decay of 0.99. The loss function is defined as

$$\mathcal{L}(\mathbf{\Theta}) = \frac{1}{T} \sum_{t=1}^{T} \|\mathbf{x}_{ref}^t - \mathbf{x}_{noisy}^t\|_2^2. \quad (5)$$

In the reconstruction stage, the trained deep denoiser is plugged into the iterative algorithm. It runs for 200 iterations with solution tolerance $10^{-8}$ and $\gamma = 1$.

The proposed method is executed on a server equipped with dual Intel Xeon Silver 4210 CPUs, 128 GB RAM, and the Nvidia Tesla T4 GPU (16 GB memory) in PyTorch 1.10.

## 3. EXPERIMENTAL RESULTS

### 3.1. Dataset

A brain dataset from fastMRI dataset [23] is used in this paper for reconstruction. It is axial T2 weighted k-space data and we select 5 cases each containing about 16 slices of size 320×320 and coil 16. All data are fully sampled, and they are retrospectively undersampled for test. ESPIRiT [24] is used to estimate coil sensitivity maps.

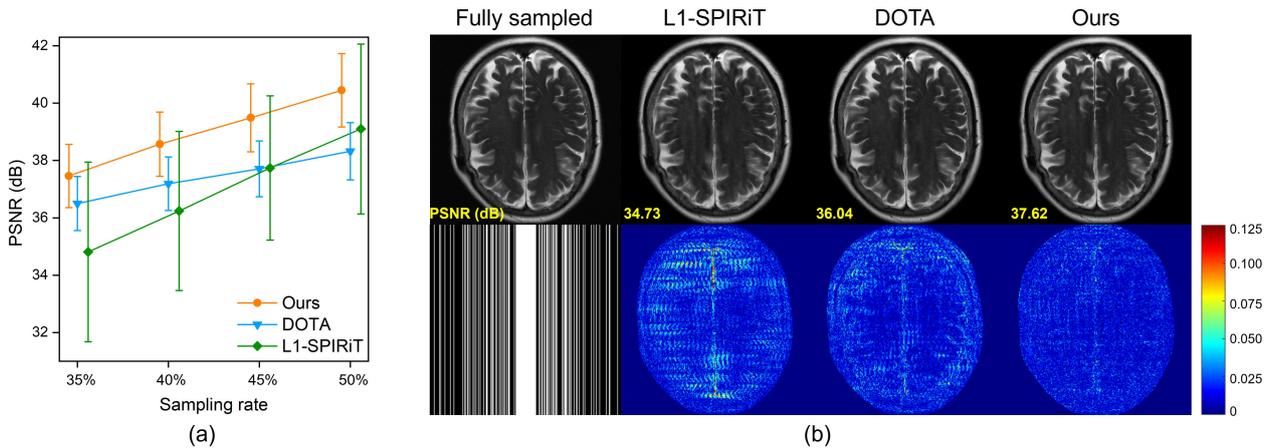

**Fig. 2. Reconstruction under 1D undersampling.** (a) is the reconstruction results of different methods under 1D Cartesian undersampling pattern with 4 sampling rates. The mean values and standard deviations of PSNR are computed over all tested cases and shifted horizontally for better visualization. (b) is the fully sampled image and a sampling pattern with sampling rate 35%, reconstructed images and the corresponding error maps.

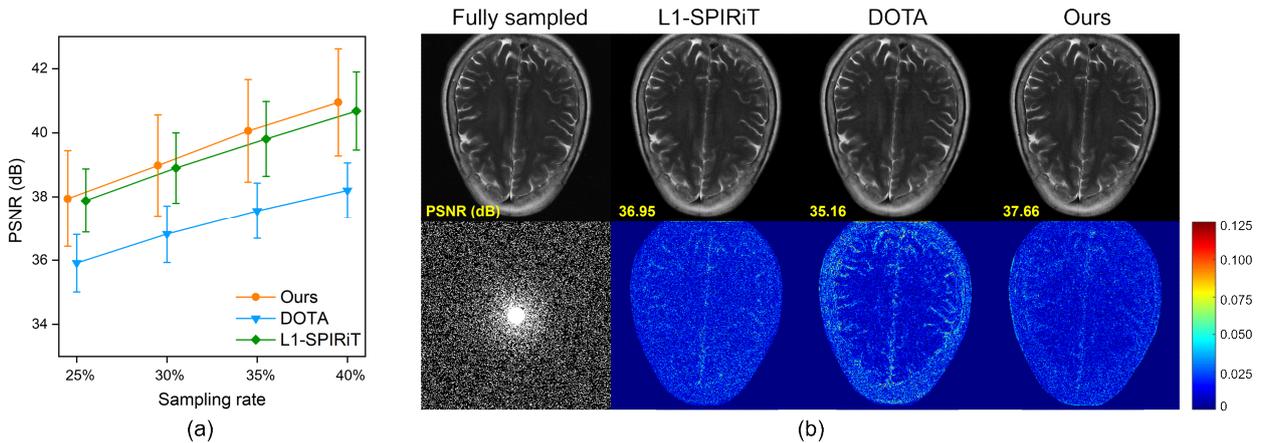

**Fig. 3. Reconstruction under 2D undersampling.** (a) is the reconstruction results of different methods under 2D random undersampling pattern with 4 sampling rates. The mean values and standard deviations of PSNR are computed over all tested cases and shifted horizontally for better visualization. (b) is the fully sampled image and a sampling pattern with sampling rate 25%, reconstructed images and the corresponding error maps.

To test robustness of reconstruction methods under different undersampling scenarios, 2 undersampling patterns are used with each having 4 sampling rates, including 1D Cartesian undersampling pattern (sampling rates are 35%, 40%, 45%, 50%) and 2D random undersampling pattern (sampling rates are 25%, 30%, 35%, 40%).

### 3.2. Evaluation criterion and compared methods

To quantitatively evaluate the reconstruction performance, we utilize the peak signal-to-noise-ratio (PSNR) as the evaluation criterion and the higher PSNR indicates the less image distortions in reconstructions.

For comparative study, conventional iterative method L1-SPIRiT [6] is used as the reconstruction baseline. We also compare the proposed method with a state-of-the-art end-to-end deep learning method DOTA [17]. DOTA is trained using 50 cases of T2-weighted brain data (each case contains about 16 slices) from fastMRI dataset [23] under 1D Cartesian undersampling pattern with sampling rate 35%. Notably, DOTA is executed according to the typical setting mentioned by the authors.

### 3.4. Reconstruction under 1D undersampling

As shown in Figure 2(a), for the tested 4 different sampling rates, our deep plug-and-play method consistently outperforms other compared methods in terms of PSNR. Differently, end-to-end trained DOTA outperforms the baseline L1-SPIRiT only when sampling rate is 35% and 40%. This may be due to the performance degradation of

DOTA when the reconstructed sampling rate deviates greatly from the trained one (i.e., 35%).

Representative images are shown in Figure 2(b). In the reconstruction under 1D Cartesian undersampling pattern with sampling rate 35%, L1-SPIRiT and DOTA yield results exhibiting some artifacts and large reconstruction errors, while the proposed method provide the images with nice artifacts suppression and details preservation.

These results imply that, the proposed method is more robust and can provide high-quality reconstructions under different sampling rates.

### 3.5. Reconstruction under 2D undersampling

In order to verify the robustness of our method to reconstructed scenarios, DOTA still uses 1D undersampling pattern for training. Thus, DOTA loses the generality for 2D undersampling discussed here. Figures 3(a)-(b) shows that our method still achieves best performance both quantitatively and visually, while other compared methods have edge artifacts and details blurring. In contrast, we can easily find that DOTA is hard to achieve satisfactory results at any sampling rate in 2D undersampling and even obviously worse than L1-SPIRiT. The reason for this phenomenon may be that it is trained in a specific 1D undersampling pattern and thus cannot generalize well to mismatched patterns.

All aforementioned results demonstrate that, the proposed method is adaptive to different sampling settings and can consistently reconstruct high-quality images on *in vivo* data.

### 4. CONCLUSION

In this work, we propose a deep plug-and-play method for undersampled MRI reconstruction, to effectively adapt to different sampling settings. We demonstrate that, a deep denoiser trained using synthetic data to remove general white Gaussian noise rather than any specific undersampling artifacts, can be directly plugged into an iterative algorithm and then work well on *in vivo* MRI reconstruction.

It may be very useful for cases where the large-scale realistic training data are unavailable and sampling settings are unknown or usually changed. Future work will include integrating advanced architectures to optimize denoiser and extending to more sampling schemes, such as non-Cartesian trajectories.

### 5. ACKNOWLEDGMENT

The authors thank Drs. Michael Lustig and Dosik Hwang for sharing their codes online. This work was supported in part by the National Natural Science Foundation of China under grants 61971361, 62122064, and 62331021, Natural Science Foundation of Fujian Province of China under grants 2023J02005, President Fund of Xiamen University under grant 20720220063, and Xiamen University Nanqiang Outstanding Talents Program.